\documentclass[conference]{IEEEtran}

\IEEEoverridecommandlockouts
\usepackage[final]{graphicx}
\usepackage{lipsum}
\usepackage{array}
\usepackage{amsmath}
\usepackage{amssymb}
\usepackage{eqparbox}
\usepackage{physics}
\usepackage[caption=false,font=footnotesize]{subfig}
\usepackage{flushend}
\usepackage{multirow}
\usepackage{xcolor}
\usepackage{tabularx}
\usepackage{algpseudocode}
\usepackage{algorithm}
\usepackage{amssymb}
\usepackage{acronym}
\usepackage[algo2e]{algorithm2e}
\usepackage{mathtools}
\usepackage[mathscr]{euscript}   
\UseRawInputEncoding
\usepackage[nocompress]{cite}

\newtheorem{example}{Example}

\usepackage{setspace}
\usepackage{hyperref}

\acrodef{AE}{autoencoder}
\acrodef{MLP}{multi-layer perceptron}
\acrodef{OFDM}{orthogonal frequency-division multiplexing}
\acrodef{KAN}{Kolmogorov-Arnold network}
\acrodef{BLER}{block error rate}
\acrodef{MLD}{maximum-likelihood decoding}
\acrodef{HDD}{hard-decision decoding}
\acrodef{SER}{symbol error rate}
\acrodef{AWGN}{additive white Gaussian noise}
\acrodef{MMSE}{minimum mean-squared error}
\acrodef{DL}{deep learning}
\acrodef{OHE}{one-hot encoded}
\acrodef{SR}{symbolic regression}
\acrodef{QPSK}{quadrature phase shift keying}
\acrodef{SiLU}{sigmoid linear unit}
\acrodef{ReLU}{rectified linear unit}

\acrodef{KAN-AE}{Kolmogorov-Arnold network-based autoencoder}
\acrodef{MLP-AE}{multi-layer perceptron-based autoencoder}
\acrodef{SR-AE}{symbolic regression autoencoder}

\def\realnumbers{\mathbb{R}}
\def\complexnumbers{\mathbb{C}}
\def\lossfunc[#1]{\mathcal{L}(#1)}
\def\Ltwonorm[#1]{||#1||_2^2}
\def\logit[#1]{l_{#1}}
\def\interval[#1][#2]{[#1, #2]}
\def\target{T_i}

\def\message{m}
\def\messagevector{\vb s_{\rm m}}
\def\messagevector{\vb s_{\rm m}}
\def\logitvector{\vb{\hat{s}_{\rm m}}}
\def\detectedmessage{\hat{m}}
\def\detectedcodeword{\hat{\vb c}}
\def\bits{k}
\def\blocks{n}
\def\rate{r}
\def\batchsize{B}

\def\channel{h}
\def\noise{w}
\def\variance[#1]{\sigma_{#1}^2}
\def\ebno{E_{\rm b}/N_0}
\def\ohevector{\vb s_{\rm e}}

\def\symbolstx{s_{\rm tx}}
\def\symbolsrx{s_{\rm rx}}
\def\estimatesymbolsrx{\hat{s}_{\rm rx}}
\def\symbolpower{\mathbb{E}[|s_{\rm tx}|^2]=1}

\def\outputvector{\textbf{s}_{\rm d}}

\def\outerfunc[#1]{\Phi_i#1}
\def\innerfunc[#1]{\phi_{i,j}(#1)}
\def\innerfuncdefault{\phi_{i,j}}
\def\layers{L}

\def\learnableFunction[#1][#2][#3][#4]{{\vb \phi}^{(#1)}_{#2,#3}\left(#4\right)}
\def\indexLayer{l}
\def\indexEdgeAtLayer[#1]{i_{#1}}
\def\dimensionAtLayer[#1]{d_{#1}}
\def\numberOfLayers{L}
\def\alayerElement[#1][#2]{a^{(#1)}_{#2}}

\def\spline[#1]{\mathtt{spline}(#1)}
\def\weightspline{w_{\rm s}}
\def\basisdegree{p}
\def\controlpoints{c}
\def\basisfunction{B}

\def\silu[#1]{\mathtt{SiLU}(#1)}
\def\weightsilu{w_{\rm b}}  
\def\epochs{n_{\rm epochs}}
\def\arbitraryPhi[#1]{\vb \Phi^{(#1)}}

\def\currsigma{\sigma^{(l)}}

\def\currPhi{\vb \Phi^{(l)}}
\def\currphi{\phi^{(l)}}

\def\currlayer{\mathbf{a}^{(l)}}
\def\prevlayer{\mathbf{a}^{(l-1)}}
\def\indlayer[#1]{a_{#1}^{(l-1)}}
\def\zerolayer[#1]{\mathbf{a}_{#1}^{(0)}}
\def\previnputs{d_{l-1}}
\def\curroutputs{d_l}

\def\errorfunction[#1]{\text{E}(#1)}
\def\defaultfunction[#1]{f(#1)}
\def\segments{N}
\def\errorsegment{e_j}
\def\scorefunction[#1]{Q[#1(x)]}

\def\scoreMLP{Q[\rm MLP (\vb x)]}
\def\scoreKAN{Q[\rm KAN (\vb x)]}

\def\Rtwo{R^2}
\def\paramone{\gamma_{\rm i}}
\def\paramtwo{\beta_{\rm i}}
\def\paramthree{\gamma_{\rm o}}
\def\paramfour{\beta_{\rm o}}

\def\functionapprox{\paramthree f_k(\paramone x+\paramtwo)+\paramfour}
\def\candidatefunction{f_k(x)}
\def\candidatefunctionslist{\{f_k(x)\}_{k=1}^K}
\def\samples{S(\phi)}

\def\phisymbolic{\phi_{\rm sym}(x)}
\def\phiapprox{\hat{\phi}_{k}(x)}
\def\phitilde{\tilde{\phi}}

\def\pruning{a_{i,j}^{(l)}}

\def\phiinc{\phi_{i,k}^{(l-1)}}
\def\phiout{\phi_{j,i}^{(l+1)}}
\def\scoreinc{I_{i}^{(l)}}
\def\scoreout{O_{i}^{(l)}}
\def\pruningthresh{\eta}

\def\encoderparams{\theta_{\rm e}}
\def\encoder[#1]{\epsilon(#1)}
\def\numberOfLayersEncoder{L_{\rm enc}}
\def\numberOfLayersDecoder{L_{\rm dec}}

\def\numberOfLayersDecoder{L_{\rm dec}}
\def\decoderparams{\theta_{\rm d}}

\IEEEoverridecommandlockouts

\begin{document}
	
	\title{KAN-AE with Non-Linearity Score and Symbolic Regression for Energy-Efficient Channel Coding}
	
	\author{Anthony Joseph Perre, Parker Huggins, and Alphan \c{S}ahin\\
		Department of Electrical Engineering, University of South Carolina, Columbia, SC, USA\\
		Email: \{aperre, parkerkh\}@email.sc.edu, asahin@mailbox.sc.edu%
		\thanks{This work has been supported by the National Science Foundation (NSF) through the award CNS-2438837.}
	}
	
	\maketitle
	
	\begin{abstract}
		In this paper, we investigate \acp{KAN-AE} with \ac{SR} for energy-efficient channel coding.  By using \ac{SR}, we convert \acp{KAN-AE} into symbolic expressions, which enables low-complexity implementation and improved energy efficiency at the radios. To further enhance the efficiency, we introduce a new non-linearity score term in the \ac{SR} process to help select lower-complexity equations when possible. Through numerical simulations, we demonstrate that \acp{KAN-AE} achieve competitive BLER performance while improving energy efficiency when paired with \ac{SR}. We score the energy efficiency of a \ac{KAN-AE} implementation using the proposed non-linearity metric and compare it to a \ac{MLP-AE}. Our experiment shows that the \ac{KAN-AE} paired with \ac{SR} uses 1.38 times less energy than the \ac{MLP-AE}, supporting that \acp{KAN-AE} are a promising choice for energy-efficient deep learning-based channel coding.
	\end{abstract}
	
	\begin{IEEEkeywords}
		autoencoder, channel coding, energy efficiency, Kolmogorov-Arnold network
	\end{IEEEkeywords}
	
	\section{Introduction}\label{Sec:1}
	
	\Ac{DL} has been demonstrated to enhance traditional signal processing techniques in modern wireless communication systems. For example, \ac{DL} approaches are used to improve channel estimation and recognize modulation type in~\cite{8640815} and ~\cite{8446021}, respectively. Furthermore, \ac{DL} has been applied to learn joint source and channel coding tasks within end-to-end \ac{OFDM} systems~\cite{8640815,8446021}. Although the application of \ac{DL} models to wireless communications shows great promise, their adoption in practical systems faces several challenges. One of the main issues lies in mobile device hardware, where constrained memory resources and CPU capabilities limit the efficacy of large \ac{DL} models at the radios\cite{8382166}. Complex models with hundreds of thousands of learnable parameters cause memory and timing issues for mobile devices, which in turn leads to increased energy and power consumption \cite{8786074}. Hence, there is a clear need for more efficient model architectures that can reduce memory usage and computational load without having to sacrifice performance. 
	
	\Acp{MLP} are a fundamental component of various \ac{DL} architectures. Recently, a novel \ac{DL} structure called \acp{KAN} have emerged as an alternative to \acp{MLP} \cite{liu2024kan}. Studies show that \acp{KAN} outperform \acp{MLP} in accuracy while requiring fewer total parameters \cite{liu2024kan}. The authors of \cite{yu2024kan} challenge some of the claims made in \cite{liu2024kan}, but confirm that \acp{KAN} outperform \acp{MLP} in symbolic formula representation under fair comparison. Recently, \acp{KAN} have seen extensive use in various domains such as physics  and time series prediction, particularly for their increased interpretability and symbolic representation capabilities \cite{liu2024kan20kolmogorovarnoldnetworks, vaca2024kolmogorov}. Of particular interest to this work is the reduced number of parameters required by \acp{KAN}, which makes them more suitable for resource-constrained environments like mobile devices, where memory and computational resources are limited. Additionally, \acp{KAN} are highly compatible with \ac{SR}, which means that the learned network can be expressed as a combination of simpler symbolic equations. These expressions can improve energy efficiency by reducing the complexity and resource demands of executing the model.
	
	In this study, we investigate the use of \acp{AE} for channel coding, which is discussed in several prior works \cite{8054694, 8445920, 9204706}. In our approach, we replace \acp{MLP} in the \ac{AE} structure with \acp{KAN}, which we find to demonstrate comparable \ac{BLER} performance while having fewer total parameters. Once the entire \ac{KAN-AE} is trained, we use \ac{SR} to derive equations representing the learned behavior. Furthermore, we introduce a non-linearity score term into the \ac{SR} process to encourage simpler equations when feasible. Our use of \ac{SR}, combined with the proposed non-linearity score term, aims to lower the energy consumption of the \ac{AE} model. By using \acp{KAN}, we show that it is possible to reduce the energy usage of certain \ac{DL} architectures without sacrificing performance, which suggests that \acp{KAN} can be a useful alternative to \acp{MLP} for specific \ac{DL} tasks within wireless communications.
	
	\textit{Organization:} The paper is organized as follows. Section~\ref{Sec:2} presents the system model and provides fundamental concepts related to \acp{KAN}. Section~\ref{Sec:3} describes the proposed \ac{KAN-AE} and  the metrics used to assess energy efficiency. Section~\ref{Sec:4} shows the \ac{BLER} performance and compares the energy efficiency of each model. Section~\ref{Sec:5} concludes the paper.
	
	\textit{Notation:} The set of real and complex numbers are denoted by $\mathbb{R}$ and $\mathbb{C}$, respectively. The complex conjugate of $z=a+jb$ is expressed as $z^{*}=a-jb$. The circularly symmetric complex normal distribution with zero mean and variance $\variance[]$ is represented as $\mathcal{CN}\left(0, \variance[]\right)$. The Hermitian of a matrix $\vb A$ is denoted by ${\vb A}^{\rm H}$. $\mathbb{E}[X]$ denotes the expected value of  $X$.

	\section{System Model} \label{Sec:2}
	In this section, we discuss preliminaries on \acp{KAN} and provide our system model on \ac{OFDM}-based \acp{AE}.
	\subsection{Kolmogorov-Arnold Networks} \label{Subsec:KAN}
	\begin{figure*}[t]
		\centering
		\includegraphics[width=7in, trim = 0.7cm 7.2cm 0.9cm 6.1cm, clip]{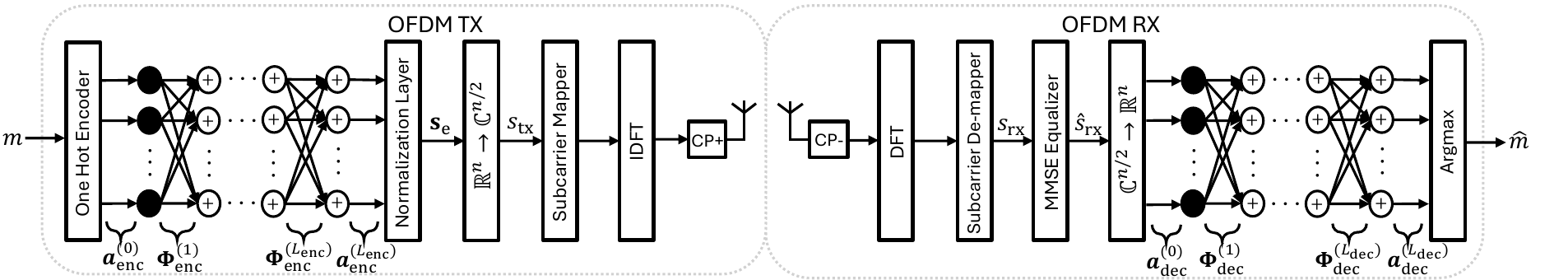}
		\caption{OFDM transmitter and receiver block diagrams with an $(\blocks,\bits)$ \ac{KAN-AE}.}
		\label{ofdm_kan_mlp}  
	\end{figure*}
	
	The structure of \acp{KAN} is inspired by the Kolmogorov-Arnold representation theorem, which establishes that any multi-variate continuous function can be expressed as the sum of several uni-variate continuous functions \cite{kolmogorov:superposition}, i.e.,
	\begin{equation} 
		\begin{aligned} \label{eq:kan} 
			\defaultfunction[x_1,x_2, \dots,x_d]&=\sum_{i=0}^{2d}\outerfunc[\left(\sum_{j=1}^{d}\innerfunc[x_j])\right]~,
		\end{aligned}
	\end{equation} where $\innerfuncdefault: [0,1]\rightarrow\realnumbers$ and $\Phi_i:\realnumbers\rightarrow\realnumbers$. The authors of \cite{liu2024kan} generalize the inner and outer sums in \eqref{eq:kan} to accommodate an arbitrary number of layers $\layers$ as
	\begin{align}
		\label{eq:kanlayers} 
		&\rm KAN (\vb x) = (\arbitraryPhi[L]_{} \circ \arbitraryPhi[L-1] \circ \dots \circ \arbitraryPhi[2] \circ \arbitraryPhi[1])(\vb x)\\&= \sum_{\indexEdgeAtLayer[\numberOfLayers-1]=1}^{\dimensionAtLayer[\numberOfLayers-1]}\learnableFunction[\numberOfLayers][\indexEdgeAtLayer[\numberOfLayers]][\indexEdgeAtLayer[\numberOfLayers-1]][\cdots{\sum_{\indexEdgeAtLayer[1]=1}^{\dimensionAtLayer[1]}\learnableFunction[2][\indexEdgeAtLayer[2]][\indexEdgeAtLayer[1]][{{\sum_{\indexEdgeAtLayer[0]=1}^{\dimensionAtLayer[0]}\learnableFunction[1][\indexEdgeAtLayer[1]][\indexEdgeAtLayer[0]][x_{\indexEdgeAtLayer[0]}]}}]}\cdots]~,\nonumber
	\end{align} 
	where 	$\learnableFunction[\indexLayer][\indexEdgeAtLayer[\indexLayer]][\indexEdgeAtLayer[\indexLayer-1]][\cdot]$ is a learnable function in the $l$th layer connecting the $\indexEdgeAtLayer[\indexLayer-1]$th input neuron to the $\indexEdgeAtLayer[\indexLayer]$th output neuron, i.e., an edge,  $\curroutputs$ is the number of neurons in the $l$th layer, $\currPhi$ contains all learnable activation functions at the $\indexLayer$th layer as
	\begin{equation} 
						\currPhi(\cdot)\triangleq
		\begin{bmatrix}
			\learnableFunction[\indexLayer][1][1][\cdot] & \ldots & \learnableFunction[\indexLayer][1][\dimensionAtLayer[\indexLayer-1]][\cdot] \\
			\vdots & \ddots & \vdots \\
			\learnableFunction[\indexLayer][\dimensionAtLayer[\indexLayer]][1][\cdot] & \ldots & \learnableFunction[\indexLayer][\dimensionAtLayer[\indexLayer]][\dimensionAtLayer[\indexLayer-1]][\cdot] \\
		\end{bmatrix},~
	\end{equation} 
	and operates on the output of the previous layer $\prevlayer\in\realnumbers^{\dimensionAtLayer[\indexLayer-1]}$, for $\forall l\in\{1,\dots,L\}$, as
	\begin{equation}
		\currlayer=\currPhi(\prevlayer)\triangleq
		\begin{bmatrix}
			\sum_{\indexEdgeAtLayer[\indexLayer-1]=1}^{\dimensionAtLayer[\indexLayer-1]}\learnableFunction[\indexLayer][\indexEdgeAtLayer[\indexLayer]][\indexEdgeAtLayer[\indexLayer-1]][{\alayerElement[\indexLayer-1][{\indexEdgeAtLayer[\indexLayer-1]}]}] \\
			\vdots\\
			\sum_{\indexEdgeAtLayer[\indexLayer-1]=1}^{\dimensionAtLayer[\indexLayer-1]}\learnableFunction[\indexLayer][\indexEdgeAtLayer[\indexLayer]][\indexEdgeAtLayer[\indexLayer-1]][{\alayerElement[\indexLayer-1][{\indexEdgeAtLayer[\indexLayer-1]}]}]
		\end{bmatrix}\nonumber~.
	\end{equation} 
	Here, an edge refers to a connection between an input and an output neuron that performs some transformation on the input. While the weights on the edges are learnable and the activation functions are fixed in \acp{MLP}, the opposite holds for \acp{KAN}.
	
	In \cite{liu2024kan}, the authors express an activation function $\phi(x)$ as a linear combination of B-splines and \ac{SiLU}, i.e.,
	\begin{align}
		 \label{eq:KANphi}
			{\vb \phi}(x) &= \weightsilu \times \silu[x] + \weightspline \times \sum_{i}\left(\controlpoints_i\times\basisfunction _i(x)\right)~,
	\end{align} where $\basisfunction_i(x)$ is a B-spline basis function, composed of piecewise polynomials of degree $\basisdegree$ and scaled by a learnable weight $\controlpoints_i$. The parameters $\weightsilu$ and $\weightspline$ are also learnable. Each B-spline is defined on a specific grid interval, which is determined by the range of input samples.
	
	\subsection{System Model} \label{Subsec:systemModel}
	Consider a single-user communication link. Let $\rate = {\bits}/{\blocks}$ be the rate of this communication link, where $\bits$ is the total number of information bits per message and $\blocks$ is the total number of channel uses. Let $\messagevector \in \realnumbers^{2^{\bits}}$ be a \ac{OHE} vector representation of the message $\message$. The encoder network  $\encoder[\messagevector]$ maps $\messagevector$ to the vector $\ohevector \in \realnumbers^{\blocks}$, which is then converted to the real and imaginary components of $\symbolstx \in \complexnumbers^{\blocks/2}$, where $\symbolpower$. We consider $\numberOfLayersEncoder$ layers at the encoder. For an \ac{MLP}-based neural network, we have
	\begin{equation}
		\begin{aligned} \label{eq:mlplayer}
			\currlayer_{\rm enc} &= \currsigma_{\rm enc}({\vb W}_{\rm enc}^{(l)}\prevlayer_{\rm enc}+{\vb b}_{\rm enc}^{(l)})~, \quad l=1,\dots, \numberOfLayersEncoder~,
		\end{aligned}
	\end{equation} where $\currsigma_{\rm enc}$ is the element-wise non-linear activation function, and ${\vb W}_{\rm enc}^{(l)} \in \realnumbers^{\curroutputs \times \previnputs}$ and ${\vb b}_{\rm enc}^{(l)} \in \realnumbers^{\curroutputs}$ are the weight matrix and bias vector, respectively. We note that for both the \ac{MLP} and \ac{KAN} models, $\zerolayer[\rm enc] = \messagevector$ and ${\vb a}_{\rm enc}^{(L_{\rm enc})} = \ohevector$. However, for a \ac{KAN}-based neural network, by re-expressing \eqref{eq:kanlayers}, we have
	\begin{equation}
		\begin{aligned} \label{eq:kanlayer}
			\currlayer_{\rm enc} &= \currPhi_{\rm enc}(\prevlayer_{\rm enc})~, \quad l=1,\dots, \numberOfLayersEncoder~, 
		\end{aligned}
	\end{equation}  
	where $\currPhi_{\rm enc}$ is $\curroutputs \times \previnputs$ function space and $\zerolayer[\rm enc] = \messagevector$. In this study, we use \eqref{eq:KANphi} to learn the activation functions in \acp{KAN}. 
	
	After encoding at the transmitter, the transmitted symbols $\symbolstx$ propagate through a communication channel, where they are distorted by the channel and by zero-mean, circularly symmetric complex \ac{AWGN}. Let $\channel$ denote the channel coefficient. A received symbol $\symbolsrx$ can then be expressed as 
	\begin{equation}
		\begin{aligned}
			\symbolsrx &= \channel\symbolstx+\noise~,
		\end{aligned}
	\end{equation} for $\noise\sim\mathcal{CN}\left(0, \variance[n]\right)$. Under an \ac{AWGN} channel, we set $\channel = 1$. For a \textit{flat-fading} Rayleigh channel, such as that observed by an \ac{OFDM} subcarrier, we instead model $\channel\sim\mathcal{CN}\left(0, 1\right)$ and assume that $h$ is known at the receiver. Then, $\symbolsrx$ are equalized using a \ac{MMSE} equalizer, which yields  $\estimatesymbolsrx = \frac{\channel^{*}\symbolsrx}{|\channel|^2+\variance[n]}$, where $\estimatesymbolsrx$ is the estimated transmitted symbol after equalization. The real and imaginary components of $\estimatesymbolsrx \in \complexnumbers^{\blocks/2}$ are then converted to a single real-valued vector $\outputvector \in \realnumbers^{\blocks}$.
	
	Let $\delta(\outputvector)$ denote the decoder network that maps $\outputvector$ to a vector $\logitvector \in \realnumbers^{2^{\bits}}$ of logarithmic odds. For $\numberOfLayersDecoder$ layers, the \ac{MLP}-based and \ac{KAN}-based decoders follow the structures in \eqref{eq:mlplayer} and \eqref{eq:kanlayer}, respectively. Furthermore, we define $\zerolayer[\rm dec] = \outputvector$ and ${\vb a}_{\rm dec}^{(L_{\rm dec})} = \logitvector$. The detected message $\detectedmessage$ is expressed as
	\begin{equation}
		\begin{aligned}
			\detectedmessage &= \arg \underset{\logitvector \in \realnumbers^{2^{\bits}}}{\max} \;\; \logitvector~.
		\end{aligned}
	\end{equation}
	
	Conventional single-layer neural networks, also known as perceptrons, can only learn linear decision boundaries, which limits their ability to capture complex non-linear relationships. While \acp{MLP} address this limitation by stacking multiple layers, a single \ac{KAN} layer is capable of modeling complex non-linear patterns due to the flexibility of its learnable activation functions. \figurename~\ref{ofdm_kan_mlp} illustrates a \ac{KAN-AE} within an end-to-end \ac{OFDM} transmitter and receiver, for a given $\blocks$ channel uses and $\bits$ bits. Although we consider an arbitrary number of \ac{KAN} layers at the transmitter and receiver, a single layer may be used in both cases. 
	
	\section{Energy-Efficient KAN-based Autoencoder} \label{Sec:3}
	In this work, we aim to reduce the number of learnable parameters in the encoder and decoder to improve energy efficiency at the transmitter and receiver, while maintaining a low \ac{BLER}. We exploit that \acp{KAN} are highly compatible with \ac{SR}, and introduce a new penalty term in the \ac{SR} process to discourage less energy-efficient symbolic expressions. The functions are heuristically scored by measuring \textit{non-linearity}, as discussed in Section~\ref{Subsec:nonlinearity} and Section~\ref{Subsec:symbolicregression}. Our approach enables scoring the energy efficiencies of \ac{KAN} and \ac{MLP} networks regardless of the implementation, as discussed in Section~\ref{Subsec:ScoringMLPandKAN}.
	
	\subsection{Quantifying Function Non-linearity}\label{Subsec:nonlinearity}
	To quantify the degree of non-linearity for a function $\defaultfunction[x]$ over an interval $\left[a, b\right]$, we propose using a piecewise linear approximation. The idea is to assess the non-linearity of $\defaultfunction[x]$ based on the minimum number of linear segments, $\segments$, required to approximate $\defaultfunction[x]$ within a specified error tolerance, $\epsilon$. The number of segments $N$ then serves as a metric of non-linearity. A larger $\segments$ indicates higher non-linearity, while a smaller $\segments$ implies that $\defaultfunction[x]$ is closer to a linear form over $\left[a, b\right]$. 
	
	To express the aforementioned metric, consider a set of uniformly spaced partition points $a_1, a_2, \ldots, a_{\segments+1}$ for $a_1=a$ and $a_{\segments+1}=b$, where $\left[a_j, a_{j+1}\right)$ is the $j$th sub-interval on which $\defaultfunction[x]$ is linearly approximated. We express the approximation error over the $j$th sub-interval $\left[a_j, a_{j+1} \right)$ as
	\begin{equation}
		\begin{aligned} \label{eq:segmenterror}
			e_j &= \int_{a_j}^{a_{j+1}} \left| \defaultfunction[x] - \psi_j(x) \right|^2 \, dx~,
		\end{aligned}
	\end{equation} where $\psi_j(x) = m_j x + k_j$ is the best-fit linear approximation of $\defaultfunction[x]$ over $\left[a_j, a_{j+1}\right)$. To obtain $\psi_j(x)$, we over-sample $\defaultfunction[x]$ in the $j$th subinterval and apply least squares regression, with $m_j$ and $k_j$ as the best-fit slope and intercept, respectively. We then measure the approximation error over all sub-intervals as
	\begin{equation}
		\begin{aligned} \label{eq:totalerror}
			\errorfunction[\segments] &= \sum_{j=1}^\segments \errorsegment~.
		\end{aligned}
	\end{equation}    
	The non-linearity measure of $\defaultfunction[x]$, denoted as $\scorefunction[f]$, is the smallest $\segments$ that satisfies $\errorfunction[\segments] < \epsilon$, i.e.,
	\begin{equation} 
		\begin{aligned} \label{eq:score}
			\scorefunction[f] &= \arg \min_N \; \errorfunction[\segments]
			\quad \text{s.t.} \quad \errorfunction[\segments] < \epsilon~.
		\end{aligned}
	\end{equation}
	If $\defaultfunction[x]$ exhibits greater non-linearity, a larger $\segments$ will be required to achieve the same approximation accuracy. Conversely, if $\defaultfunction[x]$ is more linear, a smaller $\segments$ is required. The metric for $\scorefunction[f]$ is formulated within the context of \ac{SR}. Non-linear functions are often computationally intensive and energy demanding. Therefore, by determining $\scorefunction[f]$, we can estimate the energy cost of approximating $\defaultfunction[x]$ and guide \ac{SR} towards simpler approximations when feasible.
	
	\begin{example}
		Let $\defaultfunction[x] = |5x|$ and $g(x) = \sin(5x)$ be defined on the interval $\interval[-1][1]$ and assume an error tolerance $\epsilon = 10^{-3}$. For $\defaultfunction[x]$ and $g(x)$, compute $\errorfunction[\segments]$ using \eqref{eq:segmenterror} and \eqref{eq:totalerror}. Repeat this process and increase $\segments$ each iteration until the condition in \eqref{eq:totalerror} is satisfied. Then, \eqref{eq:score} is used to determine the score for each function, yielding $\scorefunction[f]=2$ and $\scorefunction[g]=11$. This is expected, as $\sin(5x)$ is far more oscillatory on $\interval[-1][1]$ when compared to $|5x|$, thereby making it more non-linear.
	\end{example}
	
	\subsection{Symbolic Regression under Non-linearity Constraint} \label{Subsec:symbolicregression}
	Consider an activation function $\phi(x)\in\interval[a][b]\rightarrow\realnumbers$ and a finite number of candidate functions $\candidatefunctionslist$ (e.g. sin, log, exp). Obtain samples $\samples = \{\phi(x_i) \;| \;x_i \in \interval[a][b]\}$. Let $\phitilde(x)=\functionapprox$ be an approximation of $\phi(x)$ given $\paramone$, $\paramtwo$, $\paramthree$, $\paramfour$, and $\candidatefunction$. For each $\phitilde(x)$, we compute the $\Rtwo$ score
	\begin{equation}\label{eq:r2}
		\begin{aligned}
			\Rtwo[\phitilde(x)] &= 1 - \frac{\sum_{i=1}^{N} [\phi(x_i) - \phitilde(x_i)]^2}{\sum_{i=1}^{N} [\phi(x_i) - \bar{\phi}(x_i)]^2}~,
		\end{aligned}
	\end{equation}where $\bar{\phi}(x_i)=\mathbb{E}[\phi(x_i)]$. Next, we set
	\begin{equation} 
		\begin{aligned} \label{eq:phik}
			\phiapprox &= \arg \max_{\phitilde(x)} \; \Rtwo[\phitilde(x)]~,
		\end{aligned}
	\end{equation} where $\phiapprox$ is the best approximation of $\phi(x)$ for a given $\candidatefunction$. When determining the symbolic expression $\phisymbolic$ based on $\phiapprox$, we utilize \eqref{eq:score} and \eqref{eq:r2} to create a combined score term $Z[\phiapprox]$, which is given by
	\begin{equation} 
		\begin{aligned} \label{eq:zscore}
			Z[\phiapprox] &= \Rtwo[\phiapprox]+\frac{\lambda}{\scorefunction[\hat{\phi}_k]}~.
		\end{aligned}
	\end{equation} Here, $\lambda$ weights the non-linearity score term. Using the combined score in \eqref{eq:zscore}, we compute
	\begin{equation} 
		\begin{aligned} \label{eq:phisym}
			\phisymbolic &= \arg \max_{\phiapprox} \; Z[\phiapprox]~.
		\end{aligned}
	\end{equation}
	
	In this study, the parameters $\paramone$ and $\paramtwo$  maximizing $\Rtwo$ for a given $\phitilde(x)$ are determined using a grid search. Also, for each $(\paramone,\paramtwo)$ pair, $\paramthree$ and $\paramfour$ are determined using least squares regression, where $\paramthree$ and $\paramfour$ are the best-fit slope and intercept of $\samples$, respectively. The described approach builds upon \cite{liu2024kan}, with our proposed non-linearity score term added to encourage energy-efficient equations when possible. A thorough outline of the \ac{SR} procedure is presented in Algorithm \ref{alg:1}.
	
	\begin{algorithm}[t]
		\caption{Convert $\phi(x)$ to symbolic expression} \label{alg:1}
		\KwIn{$\samples$, $\candidatefunctionslist$}
		\KwOut{$\phisymbolic$}
		
		$\Rtwo_{\text{best}} = -\infty$; $Z_{\text{best}}= -\infty$; $\phiapprox=$ None; $\phisymbolic=$ None\;
		
		\For{$\candidatefunction$ \textup{in} $\candidatefunctionslist$}
		{  
			$\Rtwo_{k} = -\infty$\;
			
			\For{$(\gamma_{\rm i},\beta_{\rm i})$ \textup{in grid} $\interval[-10][10]$}
			{
				Set $\phitilde(x)=\functionapprox$. Fit $\gamma_{\rm o}$, $\beta_{\rm o}$ using linear regression with $S(\phi)$\;
				
				$\Rtwo[\phitilde(x)]$ $\leftarrow$ \eqref{eq:r2}\;
				
				\If{$\Rtwo[\phitilde(x)] > \Rtwo_{k}$}
				{
					$\phiapprox = \phitilde(x)$; $\Rtwo_{k} = \Rtwo[\phitilde(x)]$  \; 
				}
			}
			
			$Z[\phiapprox] \leftarrow$ \eqref{eq:zscore}
			
			\If{$Z[\phiapprox] > Z_{\textup{best}}$}
			{
				$\phisymbolic = \phiapprox$; $ Z_{\text{best}} = Z[\phiapprox]$\;
			}
		}
		
		\Return $\phisymbolic$
	\end{algorithm}
	
	\subsection{Scoring MLPs and KANs Based on Non-Linearity Metric}\label{Subsec:ScoringMLPandKAN}
	For a given \ac{MLP} network, the total non-linearity score combines the individual scores for linear and non-linear activations. Thus, the total score $\scoreMLP$ is expressed as
	\begin{equation} \label{eq:mlpscore}
		\begin{aligned}
			\scoreMLP &= \sum_{l=1}^{\layers}\left[\curroutputs\times \left(\previnputs+\scorefunction[\currsigma]\right)\right]~,
		\end{aligned}
	\end{equation} where $d_0$ is the first layer input size. Clearly, the choice of $\sigma^{(l)}$ in each layer affects the total score.
	
	Now consider a \ac{KAN} network, where $\currphi_{i,j}$ is the activation function that connects the $i$th input to the $j$th output in the $l$th layer. The total score $\scoreKAN$ can be determined by treating each learned activation function \textit{separately} and computing $\scorefunction[\currphi_{i,j}]$ using the method described in Section~\ref{Subsec:nonlinearity}. When computing the score for each activation function, we consider the derived \textit{symbolic expressions} for $\currphi_{i,j}$ and not the original B-spline implementation. Summing the individual scores across all activation functions in the \ac{KAN} network, we obtain
	\begin{equation} \label{eq:kanscore}
		\begin{aligned}
			\scoreKAN &= \sum_{l=1}^{\layers}\sum_{j=1}^{\curroutputs}\sum_{i=1}^{\previnputs}\left(\pruning \times \scorefunction[\currphi_{i,j}]\right)~.
		\end{aligned}
	\end{equation} Here, $\pruning$ is 0 if $\currphi_{i,j}$ is pruned and 1 otherwise. The pruning process is described in Section \ref{Subsec:pruning}. We note that for \acp{KAN}, the score of each $\currphi_{i,j}$ is determined on the grid interval of the activation function. For \acp{MLP}, this interval can be chosen based on the domain, range, and boundedness of the activation function in each hidden layer.
	
	\subsection{Details for Further Improvements}
	\subsubsection{Pruning} \label{Subsec:pruning}
	To further improve the energy efficiency of \acp{KAN}, we 
	utilize the pruning methodology in \cite{liu2024kan}.  For a \ac{KAN} with multiple layers, each neuron's importance is determined by incoming and outgoing scores
	\begin{equation} \label{eq:pruning}
		\begin{aligned}
			\scoreinc &= \max_k \left( ||\phiinc(x)||_1 \right),~ \scoreout = \max_j \left( ||\phiout(x)||_1 \right)~,
		\end{aligned}
	\end{equation}where $\phiinc(x)$ and $\phiout(x)$ represent activation functions on edges to and from the $i$th neuron in the $l$th layer. Neurons with both scores above a threshold $\pruningthresh$ are retained, and all others are pruned. For \ac{KAN} layers, we can also consider pruning individual activation functions rather than neurons. In this case, $||\phiinc(x)||_1$ is considered for all activation functions, and the edge is pruned if this value is below $\pruningthresh$. Pruning will help obtain compact closed-form expressions by removing redundant activation functions, thereby improving energy efficiency.
	
	\subsubsection{Training} To optimize BLER performance, we employ noise-scheduling, and utilize the modified cross-entropy loss function
	\begin{equation} \label{eq:loss}
		\begin{aligned}
			\mathcal{L} = -\sum_{i=1}^{2^\bits}\target\log{\left(\frac{\exp(\logit[i])}{\sum_{j=1}^{2^\bits}\exp(\logit[j])}\right)}~,
		\end{aligned}
	\end{equation} where $\target$ is the true label for the $i$th class, and $\logit[i]$ and $\logit[j]$ are the logarithmic odds for the $i$th and $j$th decoder output. The model outputs logarithmic odds directly, as there is no softmax layer at the decoder output. In this study, we use the Adam optimizer to adjust the encoder parameters $\encoderparams$ and decoder parameters $\decoderparams$ for joint training of the encoder and decoder. Algorithm \ref{alg:2} outlines this process, where $\batchsize$ is the batch size,  $\variance[\text{min}]$ and $\variance[\text{max}]$ give the noise scheduling range, and $\alpha$ is the learning rate.
	
	\begin{algorithm}[t]
		\caption{AE training with noise scheduling} \label{alg:2}
		\KwIn{$\message$, $\batchsize$, $\alpha$, $\variance[\text{max}]$, $\variance[\text{min}]$, $\epochs$}
		\KwOut{Optimized parameters $\{\encoderparams, \decoderparams\}$}
		
		$\variance[n] = \variance[\text{max}]$\;
		
		\For{$n = 1$ \textup{to} $\epochs$}{
			Sample $\{\message^{(i)}\}_{i=1}^{B}$ from $\message$\;
			
			$\symbolstx = \text{Encoder}(\message; \encoderparams)$\;
			
			$\symbolsrx = \symbolstx + \noise$, $\noise \sim \mathcal{CN}(0, \sigma^2)$\;
			
			$\hat{y} = \text{Decoder}(\symbolsrx; \decoderparams)$\;
			
			$\mathcal{L} \leftarrow \eqref{eq:loss}$\;
			
			Update $\encoderparams$ and $\decoderparams$ using $\mathcal{L}$ and $\alpha$\;
			
			$\variance[n] = \variance[\text{max}] - (n / \epochs) \times (\variance[\text{max}] - \variance[\text{min}])$\;
		}
	\end{algorithm}
	
	\section{Numerical Results}\label{Sec:4}
	For numerical experiments, we analyze (24,12) \acp{AE} in an end-to-end \ac{OFDM} system as a proof-of-concept, with plans to use larger block size in future work. For comparison, we consider \acp{MLP} with a single input, hidden, and output layer. The hidden layer uses \ac{ReLU} activation functions, while the output layer has no activations. In this study, we consider \acp{MLP} with 150 hidden layer neurons at both the encoder and decoder. For the \ac{KAN-AE}, we replace the \acp{MLP} with a single \ac{KAN} layer at both the encoder and decoder, where each learnable activation function has 5 learnable control points $\controlpoints$ and uses third-degree polynomial basis functions. To avoid removing key components of the \ac{KAN-AE}, we use a modest pruning threshold $\pruningthresh = 10^{-4}$ at the encoder and $\pruningthresh=3\times10^{-5}$ at the decoder. During the \ac{SR} procedure, we consider an error tolerance $\epsilon = 10^{-2}$ for the non-linearity score calculation described in Section~\ref{Subsec:nonlinearity}, and $\lambda = 3 \times 10^{-2}$ for the non-linearity score weight given in Section~\ref{Subsec:symbolicregression}.
	
	We create, train, and test the \ac{MLP-AE} and \ac{KAN-AE} in Python using the PyTorch machine learning library. Adam is used to train each model in \ac{AWGN} for $3 \times 10^{4}$ epochs, where each batch contains $2^{11}$ randomly selected $\message$, and the learning rate is set to $\alpha = 10^{-3}$. We use $\ebno = 0$ dB to compute $\variance[\text{min}]$ and $\ebno = 6$ dB to compute $\variance[\text{max}]$. The grid interval for each \ac{KAN} activation function is updated periodically to fit the training examples. All models are trained using an NVIDIA RTX 3070 GPU. We compare the \ac{MLP-AE} and \ac{KAN-AE} to (24,12) Golay code with \ac{MLD}. Our implementation of  \ac{MLD} for Golay code uses \ac{QPSK} as the modulation scheme, where $\symbolpower$. Let $\vb s_{\rm g} \in \complexnumbers^{\blocks/2}$ be a vector of symbols forming a modulated codeword. We implement the \ac{MLD} as 
	\begin{equation} \label{eq:mld}
		\begin{aligned}
			\detectedcodeword = \arg \max_{\vb c} \; \Re\{{\vb s}_{\rm g}^{\rm H}\vb c\}
		\end{aligned}~,
	\end{equation} where $\detectedcodeword$ is the detected modulated codeword and $\vb c \in \complexnumbers^{\blocks/2}$ is a vector containing a \ac{QPSK} modulated codeword for the Golay code. For this implementation of \ac{MLD}, the number of linear operations is $\blocks^2\times2^{\bits}$, which we use to compute a non-linearity score of $2.359296 \times 10^{6}$ for (24,12) Golay \ac{MLD}.
	
	\subsection{Block Error Rate Performance}
\begin{figure}[t]
		\centering
		\subfloat[Block error rate in AWGN.]{\includegraphics[width=2.95in]{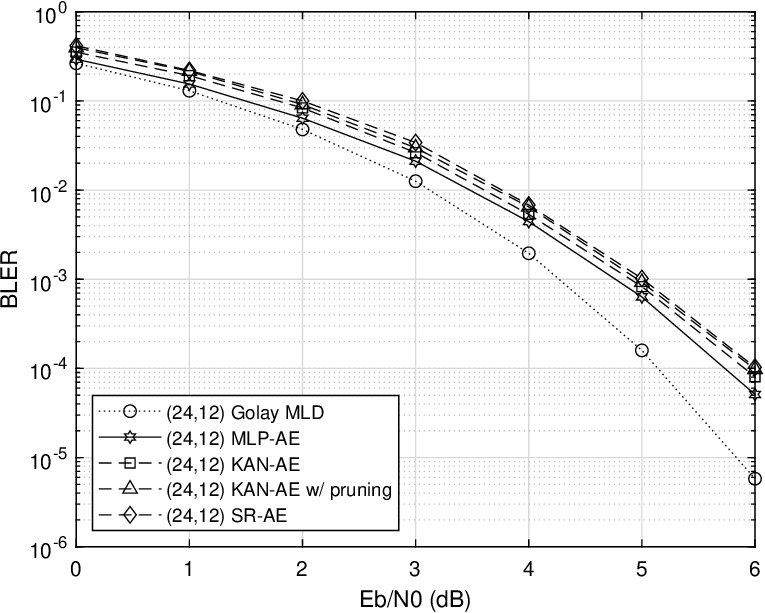}\label{BLER_AWGN}}
		\\
		\subfloat[Block error rate in Rayleigh fading channel.]{\includegraphics[width=2.95in]{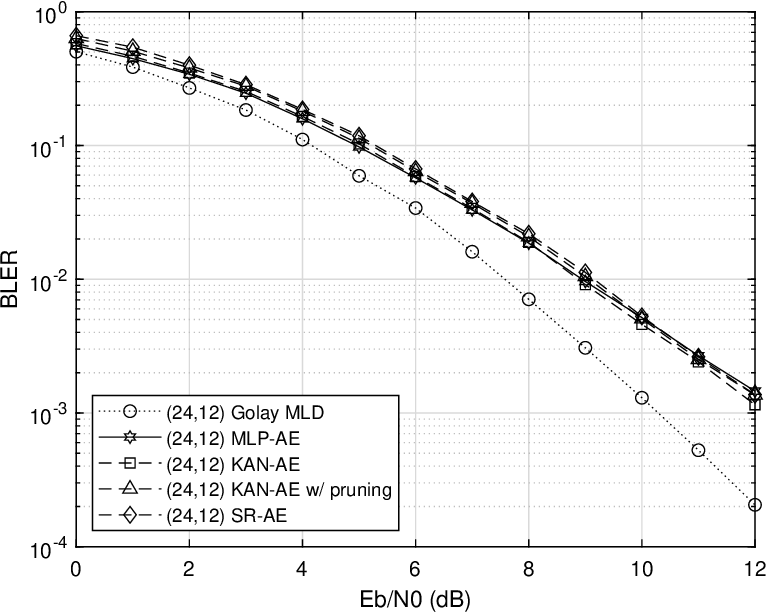}\label{BLER_REY}}	
		\caption{Performance of different (24,12) coding schemes.}	
		\label{fig:fading_error_curves}
	\end{figure}

	To characterize the \ac{BLER} performance, we perform Monte-Carlo experiments. First, we simulate the \ac{BLER} in \ac{AWGN} for the \ac{MLP-AE} and \ac{KAN-AE}, and compare it to (24,12) Golay \ac{MLD}. The \ac{BLER} curves in Fig.~\ref{BLER_AWGN} show that the \ac{KAN-AE} performs similarly to the \ac{MLP-AE} and (24,12) Golay \ac{MLD}, with (24,12) Golay \ac{MLD} slightly outperforming both. Observe that the \ac{KAN-AE} and \ac{MLP-AE} perform almost identically. Additionally, in Fig.~\ref{BLER_AWGN}, we see that pruning has a relatively minor effect on the overall \ac{BLER} performance for the \ac{KAN-AE}. We also note that the \ac{KAN}-based \ac{SR-AE} showed no loss in performance compared to the pruned model.
	
	We perform another Monte-Carlo experiment under a Rayleigh fading channel and again compare the \ac{BLER} of the \ac{MLP-AE} and \ac{KAN-AE} to (24,12) Golay \ac{MLD}. In this experiment, the symbols on each subcarrier observe \emph{flat-fading} due to the use of \ac{OFDM}, and we assume perfect channel estimation at the receiver. From Fig.~\ref{BLER_REY}, we can see that all models show comparable performance. Additionally, similar to the \ac{AWGN} channel, pruning has a slight negative effect on \ac{BLER} performance, with the \ac{SR-AE} showing nearly identical performance to the pruned model. The simulation results show that the \ac{SR-AE} performs very similar to the original \ac{KAN-AE}, thereby indicating that the model accuracy has been maintained.
	
	\begin{figure}[t]
		\centering
		\includegraphics[width=3in]{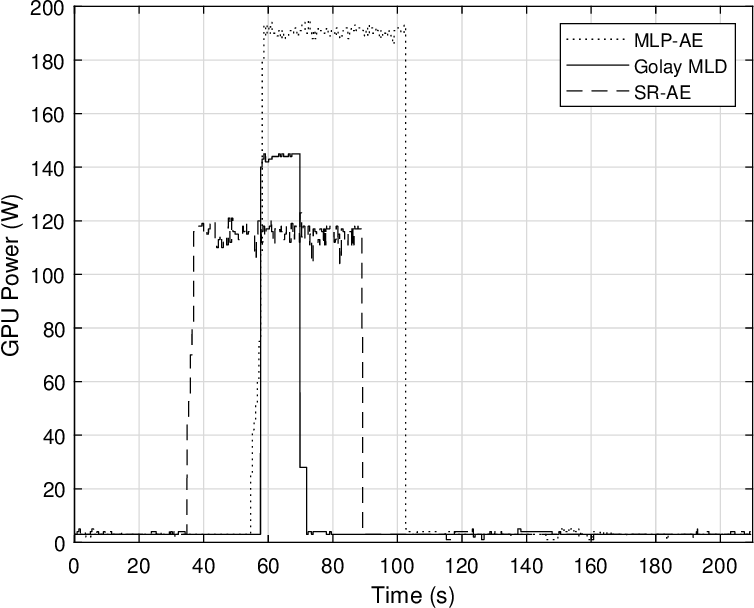}
		\caption{GPU power consumption of (24,12) coding schemes during  operation. Since this is a heuristic measurement, the non-linearity score offers a better, hardware and implementation-independent measure of the energy efficiency.}
		\label{power}  
	\end{figure}
	
	\subsection{Power and Energy Consumption}
	Another experiment is conducted where $5\times10^{3}$ messages $\message$ are processed using the \ac{MLP-AE}, \ac{SR-AE}, and Golay code, for a fixed $2.5\times10^{4}$ trials. We note that this simulation includes the encoder, channel, and decoder. We monitor the GPU power consumption during inference and compute the energy consumption for each model as the area underneath the power consumption curve. A comparison of the GPU power consumption over time for each scheme can be seen in Fig.~\ref{power}. Since an NVIDIA RTX 3070 GPU is used in this experiment, the average power draw is very large in all cases; however, in a practical system like a radio or mobile device, the power draw can be significantly reduced at the cost of evaluation speed. Also, we emphasize that the curves seen in Fig.~\ref{power} are \textit{implementation} and \textit{hardware} dependent. Therefore, it is more appropriate to use the non-linearity score to access relative energy efficiencies of each scheme since it abstracts out the implementation and hardware details. In Fig.~\ref{power}, we observe that the \ac{MLP-AE} uses approximately 1.38 times more energy as compared to the \ac{KAN}-based \ac{SR-AE}. Here, we note that \ac{MLD} for Golay code performs the best with respect to energy consumption, which can be explained by the hardware level optimizations of the PyTorch library by which it is implemented. 
	
	\begin{table}[t] 
		\centering \caption{MLP-AE vs. \ac{SR}-AE vs. Golay \ac{MLD}}
		\small
		\resizebox{\columnwidth}{!}{
			\begin{tabular}{|l|c|c|c|}
				\hline & \ac{MLP-AE} & \ac{SR-AE} & Golay with \ac{MLD}\\ \hline 
				Peak power & 195 W & 123 W & 145 W\\ \hline  Energy consumption & 9166.9 J & 6644.4 J & 2420.0 J\\ \hline
				Non-linearity score & $1.2366$e+$6$ & $6.8464$e+$5$ & $2.3592$e+$6$ \\ \hline
			\end{tabular}
		}
		\label{table:mlpvskan}
	\end{table}
	
	The peak power consumption, total energy consumption, and non-linearity score for the \ac{MLP-AE} and \ac{SR-AE} are reported in Table \ref{table:mlpvskan}. To compute the non-linearity score for the \ac{MLP-AE}, we consider the individual score for each \ac{ReLU} activation function. Since \ac{ReLU} is a piecewise linear function with $\segments=2$, a score of $2$ is assigned to each hidden layer activation. The output layer for both \acp{MLP} in the \ac{AE} have no activation, so each output layer activation function is assigned a score of $0$. Then, using \eqref{eq:mlpscore}, we calculate the score seen in Table \ref{table:mlpvskan}. Next, consider the \ac{KAN-AE}, which is pruned and converted to symbolic expressions. Each activation function is considered on its grid interval, which is $\left[0, 1\right]$ for those in the encoder and $\left[-2.2, 2.2\right]$ for those in the decoder. Using \eqref{eq:kanscore}, we calculate the score for the \ac{SR-AE} seen in Table \ref{table:mlpvskan}.
	
	\section{Concluding Remarks} \label{Sec:5}
	This study demonstrates that \acp{KAN} can provide advantages over \acp{MLP} in terms of energy efficiency and model size for channel coding tasks. The ability to derive symbolic expressions from \acp{KAN} allows for simplified, low-complexity equations, which reduces the computational burden during inference. To obtain simpler symbolic expressions, we propose a non-linearity metric, which we then use to score different symbolic expressions and eliminate unnecessary, highly non-linear activation functions during the \ac{SR} process. Our results show that \acp{KAN-AE} perform similarly to \acp{MLP-AE} in \ac{AWGN} and flat-fading Rayleigh channels, while achieving reduced energy consumption when combined with the proposed \ac{SR} method. This makes \acp{KAN} a promising option for integrating \ac{DL} models into energy-constrained devices within practical communication systems. Future work will focus on refining the non-linearity metric, considering larger codes, and performing a sensitivity analysis to examine how the score term impacts the \ac{SR} process.
	
	\bibliographystyle{IEEEtran}
	\bibliography{IEEEabrv,references}

\begin{thebibliography}{10}
\providecommand{\url}[1]{#1}
\csname url@samestyle\endcsname
\providecommand{\newblock}{\relax}
\providecommand{\bibinfo}[2]{#2}
\providecommand{\BIBentrySTDinterwordspacing}{\spaceskip=0pt\relax}
\providecommand{\BIBentryALTinterwordstretchfactor}{4}
\providecommand{\BIBentryALTinterwordspacing}{\spaceskip=\fontdimen2\font plus
\BIBentryALTinterwordstretchfactor\fontdimen3\font minus
  \fontdimen4\font\relax}
\providecommand{\BIBforeignlanguage}[2]{{%
\expandafter\ifx\csname l@#1\endcsname\relax
\typeout{** WARNING: IEEEtran.bst: No hyphenation pattern has been}%
\typeout{** loaded for the language `#1'. Using the pattern for}%
\typeout{** the default language instead.}%
\else
\language=\csname l@#1\endcsname
\fi
#2}}
\providecommand{\BIBdecl}{\relax}
\BIBdecl

\bibitem{8640815}
M.~Soltani, V.~Pourahmadi, A.~Mirzaei, and H.~Sheikhzadeh, ``Deep
  learning-based channel estimation,'' \emph{IEEE Communications Letters},
  vol.~23, no.~4, pp. 652--655, 2019.

\bibitem{8446021}
M.~Zhang, Y.~Zeng, Z.~Han, and Y.~Gong, ``Automatic modulation recognition
  using deep learning architectures,'' in \emph{Proc. IEEE International
  Workshop on Signal Processing Advances in Wireless Communications (SPAWC)},
  2018, pp. 1--5.

\bibitem{8382166}
Q.~Mao, F.~Hu, and Q.~Hao, ``Deep learning for intelligent wireless networks: A
  comprehensive survey,'' \emph{IEEE Communications Surveys \& Tutorials},
  vol.~20, no.~4, pp. 2595--2621, 2018.

\bibitem{8786074}
H.~Huang, S.~Guo, G.~Gui, Z.~Yang, J.~Zhang, H.~Sari, and F.~Adachi, ``Deep
  learning for physical-layer {5G} wireless techniques: Opportunities,
  challenges and solutions,'' \emph{IEEE Wireless Communications}, vol.~27,
  no.~1, pp. 214--222, 2020.

\bibitem{liu2024kan}
Z.~Liu, Y.~Wang, S.~Vaidya, F.~Ruehle, J.~Halverson, M.~Solja{\v{c}}i{\'c},
  T.~Y. Hou, and M.~Tegmark, ``{KAN}: {Kolmogorov-Arnold} networks,''
  \emph{arXiv preprint arXiv:2404.19756}, 2024.

\bibitem{yu2024kan}
R.~Yu, W.~Yu, and X.~Wang, ``{KAN} or {MLP}: A fairer comparison,'' \emph{arXiv
  preprint arXiv:2407.16674}, 2024.

\bibitem{liu2024kan20kolmogorovarnoldnetworks}
\BIBentryALTinterwordspacing
Z.~Liu, P.~Ma, Y.~Wang, W.~Matusik, and M.~Tegmark, ``{KAN} 2.0:
  {Kolmogorov-Arnold} networks meet science,'' 2024. [Online]. Available:
  \url{https://arxiv.org/abs/2408.10205}
\BIBentrySTDinterwordspacing

\bibitem{vaca2024kolmogorov}
C.~J. Vaca-Rubio, L.~Blanco, R.~Pereira, and M.~Caus, ``{Kolmogorov-Arnold}
  networks ({KANs}) for time series analysis,'' \emph{arXiv preprint
  arXiv:2405.08790}, 2024.

\bibitem{8054694}
T.~O’Shea and J.~Hoydis, ``An introduction to deep learning for the physical
  layer,'' \emph{IEEE Transactions on Cognitive Communications and Networking},
  vol.~3, no.~4, pp. 563--575, 2017.

\bibitem{8445920}
A.~Felix, S.~Cammerer, S.~Dörner, J.~Hoydis, and S.~Ten~Brink,
  ``{OFDM}-autoencoder for end-to-end learning of communications systems,'' in
  \emph{Proc. IEEE International Workshop on Signal Processing Advances in
  Wireless Communications (SPAWC)}, 2018, pp. 1--5.

\bibitem{9204706}
D.~Wu, M.~Nekovee, and Y.~Wang, ``Deep learning-based autoencoder for m-user
  wireless interference channel physical layer design,'' \emph{IEEE Access},
  vol.~8, pp. 174\,679--174\,691, 2020.

\bibitem{kolmogorov:superposition}
A.~K. Kolmogorov, ``On the representation of continuous functions of several
  variables by superposition of continuous functions of one variable and
  addition,'' \emph{Doklady Akademii Nauk SSSR}, vol. 114, pp. 369--373, 1957.

\end{thebibliography}
	
\end{document}